\documentclass[twocolumn,superscriptaddress,showkeys]{revtex4}
\usepackage{amssymb,epsf,epsfig,amsmath,color}
\usepackage{multirow}
\usepackage{accents}
\newcommand*{\fancybar}{\scalebox{.4}{(}\raisebox{-1.7pt}{-}\scalebox{.4}{)}}
\newcommand*{\brabar}[1]{\accentset{\fancybar}{#1}}
\newcommand{\lsim}{
\mathrel{\hbox{\rlap{\hbox{\lower4pt\hbox{$\sim$}}}\hbox{$<$}}}}
\newcommand{\gsim}{
\mathrel{\hbox{\rlap{\hbox{\lower4pt\hbox{$\sim$}}}\hbox{$>$}}}}

\def\D0{D\O }
\def\theabstract{We have recently seen new upper bounds for $B^0_s\to \mu^+\mu^-$, 
a key decay to search for physics beyond the Standard Model. 
Furthermore a non-vanishing decay width difference $\Delta\Gamma_s$ of the $B_s$ system 
has been measured. We show that $\Delta\Gamma_s$ affects the extraction of the 
$B^0_s\to \mu^+\mu^-$ branching ratio and the resulting constraints on the New Physics 
parameter space, and give formulae for including this effect. Moreover, we point out that 
$\Delta\Gamma_s$ provides a new observable, the effective $B^0_s\to \mu^+\mu^-$ lifetime 
$\tau_{\mu^+\mu^-}$, which offers a theoretically clean probe for New Physics searches that is 
complementary to the branching ratio. Should the $B^0_s\to \mu^+\mu^-$ branching ratio
agree with the Standard Model, the measurement of $\tau_{\mu^+\mu^-}$, which appears 
feasible at upgrades of the LHC experiments, may still reveal large New Physics effects.}
\begin{document}
\begin{titlepage}
\vspace*{1.7truecm}
\begin{flushright}
Nikhef-2012-006
\end{flushright}

\vspace{1.6truecm}

\begin{center}
\boldmath
{\Large{\bf Probing New Physics via the $B^0_s\to \mu^+\mu^-$ Effective Lifetime}}
\unboldmath
\end{center}

\vspace{1.2truecm}

\begin{center}
{\bf Kristof De Bruyn\,${}^a$, Robert Fleischer\,${}^{a,b}$, Robert Knegjens\,${}^a$, \\
Patrick Koppenburg\,${}^a$, Marcel Merk\,${}^{a,b}$, Antonio Pellegrino\,${}^a$,
Niels Tuning\,${}^a$}

\vspace{0.5truecm}

${}^a${\sl Nikhef, Science Park 105, NL-1098 XG Amsterdam, The Netherlands}

${}^b${\sl  Department of Physics and Astronomy, Vrije Universiteit Amsterdam,\\
NL-1081 HV Amsterdam, The Netherlands}

\end{center}

\vspace*{2.7cm}

\begin{center}
\large{\bf Abstract}\\

\vspace*{0.6truecm}

\begin{tabular}{p{13.5truecm}}
\theabstract
\end{tabular}

\end{center}

\vspace*{3.7truecm}

\vfill

\noindent
April 2012

\end{titlepage}

\newpage
\thispagestyle{empty}
\mbox{}

\newpage
\thispagestyle{empty}
\mbox{}

\rule{0cm}{23cm}

\newpage
\thispagestyle{empty}
\mbox{}

\setcounter{page}{0}

\preprint{Nikhef-2012-nnn}

\date{\today}

\title{\boldmath Probing New Physics via the $B^0_s\to \mu^+\mu^-$ Effective Lifetime
\unboldmath}

\author{Kristof De Bruyn}
\affiliation{Nikhef, Science Park 105, NL-1098 XG Amsterdam, The Netherlands}

\author{Robert Fleischer}
\affiliation{Nikhef, Science Park 105, NL-1098 XG Amsterdam, The Netherlands}
\affiliation{Department of Physics and Astronomy, Vrije Universiteit Amsterdam,
NL-1081 HV Amsterdam, The Netherlands}

\author{Robert Knegjens}
\affiliation{Nikhef, Science Park 105, NL-1098 XG Amsterdam, The Netherlands}

\author{Patrick Koppenburg}
\affiliation{Nikhef, Science Park 105, NL-1098 XG Amsterdam, The Netherlands}

\author{Marcel Merk}
\affiliation{Nikhef, Science Park 105, NL-1098 XG Amsterdam, The Netherlands}
\affiliation{Department of Physics and Astronomy, Vrije Universiteit Amsterdam,
NL-1081 HV Amsterdam, The Netherlands}

\author{Antonio Pellegrino}
\affiliation{Nikhef, Science Park 105, NL-1098 XG Amsterdam, The Netherlands}

\author{Niels Tuning}
\affiliation{Nikhef, Science Park 105, NL-1098 XG Amsterdam, The Netherlands}

\begin{abstract}
\vspace{0.2cm}\noindent
\theabstract
\end{abstract}


\maketitle

\section{Introduction}
Thanks to the Large Hadron Collider (LHC) at CERN we have entered a new era of 
particle physics. One of the most promising processes for probing the 
quark-flavor sector of the Standard Model (SM) is the rare decay $B^0_s\to \mu^+\mu^-$. 
In the SM, it originates only from box and penguin topologies, and the CP-averaged 
branching ratio is predicted to be \cite{buras}
\begin{equation}\label{BR-SM}
\mbox{BR}(B_s\to \mu^+\mu^-)_{\rm SM}=(3.2\pm0.2)\times 10^{-9},
\end{equation}
where the error is fully dominated by non-perturbative QCD effects determined 
through lattice studies. The most stringent 
experimental upper bound on this branching ratio is given by 
$\mbox{BR}(B_s\to \mu^+\mu^-)< 4.5 \times 10^{-9}$ at the 95\% confidence
level (C.L.) \cite{LHCb-Bsmumu}.  

In the presence of ``New Physics" (NP), there may be additional contributions through 
new particles in the loops or new contributions at the tree level, which are forbidden in 
the SM (see Ref.~\cite{buras} and references therein). 

A key feature of the $B_s$-meson system is $B^0_s$--$\bar B^0_s$ mixing. This
quantum mechanical effect gives rise to time-dependent oscillations between
the $B^0_s$ and $\bar B^0_s$ states. In contrast to the $B_d$ system, we
expect a sizable difference 
$\Delta\Gamma_s\equiv \Gamma_{\rm L}^{(s)} - \Gamma_{\rm H}^{(s)}$
between the decay widths of the light and heavy $B_s$ mass eigenstates \cite{LN}. 

Performing a time-dependent analysis of $B^0_s \to J/\psi \phi$,  the LHCb collaboration 
has recently reported $\Delta\Gamma_s=(0.116\pm0.019)\:\mbox{ps}^{-1}$
\cite{LHCb-Mor-12}, which represents the current most precise measurement of this observable.

As we pointed out in Ref.~\cite{BR-Bs}, the sizable  $\Delta\Gamma_s$ complicates the 
extraction of the branching ratios of $B_s$-meson decays, leading to systematic biases 
as large as ${\cal O}(10\%)$ that depend on the dynamics of the decay at hand. 

In the case of the $B^0_s\to \mu^+\mu^-$ channel, the comparison of the experimentally 
measured branching ratio with the theoretical prediction (\ref{BR-SM}) is also affected 
by this effect, which has so far been neglected in the literature. 

It can be included through a measurement of the effective $B^0_s\to \mu^+\mu^-$ lifetime. 
As $B^0_s\to \mu^+\mu^-$ 
is a rare decay, it turns out that this observable offers another sensitive probe for NP
that is {\it theoretically clean} and complementary to the branching ratio. 

\boldmath
\section{The General $B_s\to \mu^+\mu^-$ Amplitudes}
\unboldmath
The general low-energy effective Hamiltonian for the $\bar B^0_s\to \mu^+\mu^-$ 
decay can be written as 
\begin{displaymath}
{\cal H}_{\rm eff}=-\frac{G_{\rm F}}{\sqrt{2}\pi} V_{ts}^\ast V_{tb} \alpha
\bigl[C_{10} O_{10} + C_{S} O_S + C_P O_P
\end{displaymath}
\vspace*{-0.5truecm}
\begin{equation}\label{Heff}
+ C_{10}' O_{10}' + C_{S}' O_S' + C_P' O_P' \bigr].
\end{equation}
Here $G_{\rm F}$ is Fermi's constant, the $V_{qq'}$ are elements of the 
Cabibbo--Kobayashi--Maskawa (CKM) matrix, $\alpha$ is the QED fine structure constant, 
the $C_i$,  $C_i'$ are Wilson coefficients encoding the short-distance physics, 
while the 
\begin{eqnarray}
O_{10}&=&(\bar s \gamma_\mu P_L b) (\bar\ell\gamma^\mu \gamma_5\ell) \nonumber\\
O_S&=&m_b (\bar s P_R b)(\bar \ell \ell) \\
O_P&=&m_b (\bar s P_R b)(\bar \ell \gamma_5 \ell) \nonumber
\end{eqnarray}
are four-fermion operators with $P_{L,R}\equiv(1\mp\gamma_5)/2$ and $m_b$ is 
the $b$-quark mass. 
The $O'_i$  are obtained from the $O_i$ by making the replacements $P_L \leftrightarrow P_R$.
Only operators resulting in non-vanishing contributions to $\bar B^0_s\to \mu^+\mu^-$
are included in (\ref{Heff}). In particular the matrix elements of  operators 
involving the $\bar \ell\gamma^\mu\ell$ vector current vanish.

This notation is similar to Ref.~\cite{APS}, where a model-independent analysis of 
NP effects in $b\to s$ transitions was performed. In the SM, as assumed in 
(\ref{BR-SM}), only $C_{10}$ is non-vanishing and given by the real coefficient 
$C_{10}^{\rm SM}$. An outstanding feature of $\bar B^0_s\to \mu^+\mu^-$ is 
the sensitivity to (pseudo-)scalar lepton densities, as described by the $O_{(P)S}$ and 
$O_{(P)S}'$ operators. Their Wilson coefficients are still largely unconstrained and leave 
ample space for  NP. 

The hadronic sector of the leptonic 
$\bar B^0_s\to \mu^+\mu^-$ decay can be expressed in terms of a single,
non-perturbative parameter, the $B_s$-meson decay constant $f_{Bs}$ \cite{buras}.

For the discussion of the observables in Section~\ref{sec:obs}, we go to the
rest frame of the decaying $\bar B^0_s$ meson and distinguish between the
$\mu^+_{\rm L}\mu^-_{\rm L}$ and $\mu^+_{\rm R}\mu^-_{\rm R}$ helicity configurations, 
which we denote as $\mu_\lambda^+\mu_\lambda^-$ with $\lambda={\rm L, R}$. 
In this notation, $\mu_{\rm L}^+\mu_{\rm L}^-$ and $\mu_{\rm R}^+\mu_{\rm R}^-$ are 
related to each other through a CP transformation: 
\begin{equation}\label{CP-1}
|(\mu_{\rm L}^+\mu_{\rm L}^-)_{\rm CP}\rangle \equiv ({\cal CP})|\mu_{\rm L}^+\mu_{\rm L}^-\rangle=
e^{i\phi_{\rm CP}(\mu\mu)}|\mu_{\rm R}^+\mu_{\rm R}^-\rangle,
\end{equation}
where $e^{i\phi_{\rm CP}(\mu\mu)}$ is convention-dependent. We then obtain
\begin{displaymath}
A(\bar B^0_s \to \mu_\lambda^+\mu_\lambda^-)=\langle \mu_\lambda^-\mu_\lambda^+|
{\cal H}_{\rm eff}| \bar B^0_s \rangle = -\frac{G_{\rm F}}{\sqrt{2}\pi} V_{ts}^\ast V_{tb} \alpha
\end{displaymath}
\vspace*{-0.7truecm}
\begin{equation}\label{ampl}
\times f_{B_s} M_{B_s}   m_\mu C_{10}^{\rm SM} e^{i\phi_{\rm CP}(\mu\mu)(1-\eta_\lambda)/2}
\left [\eta_\lambda P +  S    \right],
\end{equation}
where $M_{B_s}$ is the $B_s$ mass, $\eta_{\rm L}=+1$ and  $\eta_{\rm R}=-1$, and
\begin{equation}\label{P-expr}
P\equiv \frac{C_{10}-C_{10}'}{C_{10}^{\rm SM}}+\frac{M_{B_s} ^2}{2 m_\mu}
\left(\frac{m_b}{m_b+m_s}\right)\left(\frac{C_P-C_P'}{C_{10}^{\rm SM}}\right)
\end{equation}
\begin{equation}\label{S-expr}
S\equiv \sqrt{1-4\frac{m_\mu^2}{M_{B_s}^2}}
\frac{M_{B_s} ^2}{2 m_\mu}\left(\frac{m_b}{m_b+m_s}\right)
\left(\frac{C_S-C_S'}{C_{10}^{\rm SM}}\right).
\end{equation}
The $P\equiv |P|e^{i\varphi_P}$ and $S\equiv |S|e^{i\varphi_S}$ carry, in general, non-trivial
CP-violating phases $\varphi_P$ and $\varphi_S$. However, in the SM, we simply have 
$P=1$ and $S=0$ (see also Ref.~\cite{APS}). The $\phi_{\rm CP}(\mu\mu)$ factor in
(\ref{ampl}) originates from using the operator relation 
$({\cal CP})^\dagger ({\cal CP})=\hat 1$ and (\ref{CP-1}) in the leptonic parts of the 
four-fermion operators.

\boldmath
\section{The $B_s\to \mu^+\mu^-$ Observables}\label{sec:obs}
\unboldmath
For the observables discussed below we need the
\begin{equation}
A(B^0_s \to \mu_\lambda^+\mu_\lambda^-)=\langle \mu_\lambda^-\mu_\lambda^+|
{\cal H}_{\rm eff}^\dagger | B^0_s \rangle 
\end{equation}
amplitude. Inserting again $({\cal CP})^\dagger ({\cal CP})=\hat 1$ into the matrix elements of
the four-fermion operators and using both (\ref{CP-1}) and 
$({\cal CP})|B^0_s\rangle=e^{i\phi_{\rm CP}(B_s)}|\bar B^0_s\rangle$, we obtain
\begin{displaymath}
A(B^0_s \to \mu_\lambda^+\mu_\lambda^-)= -\frac{G_{\rm F}}{\sqrt{2}\pi} 
V_{ts} V_{tb}^\ast \alpha f_{B_s} M_{B_s}  m_\mu C_{10}^{\rm SM}
\end{displaymath}
\vspace*{-0.7truecm}
\begin{equation}\label{ampl-CP}
\times \, 
e^{i[\phi_{\rm CP}(B_s)+\phi_{\rm CP}(\mu\mu)(1-\eta_\lambda)/2]}
\left [-\eta_\lambda P^\ast +  S^\ast    \right],
\end{equation}
which should be compared with (\ref{ampl}). We observe that
\begin{equation}
|A(B^0_s \to \mu_{\rm L, R}^+\mu_{\rm L, R}^-)| = 
|A(\bar B^0_s \to \mu_{\rm R, L}^+\mu_{\rm R, L}^-)|.
\end{equation}

Following the formalism to describe $B^0_s$--$\bar B^0_s$ mixing discussed in 
Ref.~\cite{RF-habil}, we consider the observable
\begin{displaymath}
\xi_\lambda\equiv - e^{-i\phi_s}\left[ e^{i\phi_{\rm CP}(B_s)}
\frac{A(\bar B^0_s\to \mu_\lambda^+\mu_\lambda^-)}{A(B^0_s\to \mu_\lambda^+\mu_\lambda^-)}
\right]
\end{displaymath}
\vspace*{-0.4truecm}
\begin{equation}
=-\left[\frac{+\eta_\lambda P \,+\,  S}{-\eta_\lambda P^\ast + S^\ast }\right].
\end{equation}
Here we have taken into account that the $B^0_s$--$\bar B^0_s$ mixing phase
$\phi_s\equiv2\mbox{arg}(V_{ts}^\ast V_{tb})$ is cancelled by the CKM factors
in (\ref{ampl}) and (\ref{ampl-CP}), and that the convention-dependent phase $\phi_{\rm CP}(B_s)$
is cancelled through (\ref{ampl-CP}), whereas  $\phi_{\rm CP}(\mu\mu)$ simply cancels 
in the amplitude ratio. We notice the relation
\begin{equation}\label{xiLxiR}
\xi_{\rm L}^{\phantom{\ast}} \xi_{\rm R}^{\ast} = \xi_{\rm R}^{\phantom{\ast}} \xi_{\rm L}^{\ast} =1.
\end{equation}

The observables $\xi_\lambda$ contain all the information for calculating the 
time-dependent rate asymmetries \cite{RF-habil}:
\begin{displaymath}
\frac{\Gamma(B^0_s(t)\to \mu_\lambda^+\mu^-_\lambda)-
\Gamma(\bar B^0_s(t)\to \mu_\lambda^+
\mu^-_\lambda)}{\Gamma(B^0_s(t)\to \mu_\lambda^+\mu^-_\lambda)+
\Gamma(\bar B^0_s(t)\to \mu_\lambda^+\mu^-_\lambda)}
\end{displaymath}
\vspace*{-0.3truecm}
\begin{equation}\label{asym-1}
=\frac{C_\lambda\cos(\Delta M_st)+S_\lambda\sin(\Delta M_st)}{\cosh(y_st/\tau_{B_s}) + 
{\cal A}_{\Delta\Gamma}^\lambda \sinh(y_st/\tau_{B_s})}.
\end{equation}
Here $\Delta M_s$ is the mass difference of the heavy and light $B_s$ mass eigenstates, and 
\begin{equation}\label{ys-LHCb}
	y_s \equiv \tau_{B_s} \Delta\Gamma_s/2= 0.088 \pm 0.014,
\end{equation}
where $\tau_{B_s}$ is the $B_s$ mean lifetime; the numerical value corresponds
to the results of Ref.~\cite{LHCb-Mor-12}. CP asymmetries of this kind were
considered for $B_{s,d}\to\ell^+\ell^-$ decays (neglecting $\Delta\Gamma_s$) in various
NP scenarios in Refs.~\cite{HL,DP,CKWW}.

The observables entering (\ref{asym-1})
are given as follows:
\begin{equation}\label{C-lam}
C_\lambda\equiv\frac{1-|\xi_\lambda|^2}{1+|\xi_\lambda|^2}
=-\eta_\lambda\left[\frac{2|PS|\cos(\varphi_P-\varphi_S)}{|P|^2+|S|^2}  \right]
\end{equation}
\begin{equation}\label{S-lam}
S_\lambda\equiv \frac{2\,\mbox{Im}\,\xi_\lambda}{1+|\xi_\lambda|^2}
=\frac{|P|^2\sin 2\varphi_P-|S|^2\sin 2\varphi_S}{|P|^2+|S|^2}
\end{equation}
\begin{equation}\label{ADG-lam}
{\cal A}_{\Delta\Gamma}^\lambda\equiv \frac{2\,\mbox{Re}\,\xi_\lambda}{1+|\xi_\lambda|^2}
=\frac{|P|^2\cos 2\varphi_P-|S|^2\cos 2\varphi_S}{|P|^2+|S|^2}.
\end{equation}
It should be emphasized that due to (\ref{xiLxiR}) ${\cal S}_{\rm CP} \equiv S_\lambda$ and 
${\cal A}_{\Delta\Gamma}\equiv {\cal A}_{\Delta\Gamma}^\lambda$ do not depend on the
helicity $\lambda$ of the muons and are {\it theoretically clean} observables.

Since it is difficult to measure the muon helicity, we consider the rates
\begin{equation}
\Gamma(\brabar{B}_s^0(t)\to \mu^+\mu^-)\equiv \sum_{\lambda={\rm L,R}}
\Gamma(\brabar{B}_s^0(t)\to \mu^+_\lambda \mu^-_\lambda),
\end{equation}
and obtain then the CP-violating rate asymmetry
\begin{displaymath}
\frac{\Gamma(B^0_s(t)\to \mu^+\mu^-)-
\Gamma(\bar B^0_s(t)\to \mu^+\mu^-)}{\Gamma(B^0_s(t)\to \mu^+\mu^-)+
\Gamma(\bar B^0_s(t)\to \mu^+\mu^-)}
\end{displaymath}
\vspace*{-0.3truecm}
\begin{equation}\label{asym-2}
=\frac{{\cal S}_{\rm CP}\sin(\Delta M_st)}{\cosh(y_st/ \tau_{B_s}) + 
{\cal A}_{\Delta\Gamma} \sinh(y_st/ \tau_{B_s})},
\end{equation}
where the $C_\lambda$ terms (\ref{C-lam}) cancel because of the $\eta_\lambda$ factor.

It would be most interesting to measure (\ref{asym-2}) since a non-zero value immediately 
signaled CP-violating NP phases. Unfortunately, this is challenging in 
view of the tiny branching ratio and as tagging, distinguishing between initially present 
$B^0_s$ and $\bar B^0_s$ mesons, and time information are required. An expression 
analogous to (\ref{asym-2}) holds also for $B_d\to\mu^+\mu^-$ decays.

In practice, the branching ratio
\begin{equation}\label{defBrExp}
	{\rm BR}\left(B_s \to \mu^+\mu^-\right)_{\rm exp} 
	\equiv \frac{1}{2}\int_0^\infty \langle \Gamma(B_s(t)\to \mu^+\mu^-)\rangle\, dt
\end{equation}
is the first measurement, where the ``untagged" rate
\begin{displaymath}
\langle \Gamma(B_s(t)\to f)\rangle\equiv
\Gamma(B^0_s(t)\to f)+ \Gamma(\bar B^0_s(t)\to f)
\end{displaymath}
\vspace*{-0.5truecm}
\begin{equation}\label{untagged}
\propto e^{-t/\tau_{B_s}}\bigl[\cosh(y_st/ \tau_{B_s})+  {\cal A}_{\Delta\Gamma}
\sinh(y_st/ \tau_{B_s})\bigr]
\end{equation}
is introduced \cite{BR-Bs,DFN}. The branching ratio (\ref{defBrExp}) is extracted  
ignoring tagging and time information. 
As shown in Ref.~\cite{BR-Bs}, due to the sizable width difference,
the experimental value (\ref{defBrExp}) is related to the theoretical value 
(calculated in the literature, see, e.g., Refs.\cite{buras,APS}) through 
\begin{equation}\label{BRratio}
	{\rm BR}(B_s \to \mu^+\mu^-)= 
	\left[\frac{1-y_s^2}{1 + {\cal A}_{\Delta\Gamma}\, y_s}\right] 
	{\rm BR}(B_s \to \mu^+\mu^-)_{\rm exp},
\end{equation}
where
\begin{equation}
\frac{\mbox{BR}(B_s\to\mu^+\mu^-)}{\mbox{BR}(B_s\to\mu^+\mu^-)_{\rm SM}}=
|P|^2+|S|^2.
\end{equation}
The $y_s$ terms in (\ref{BRratio}) were so far not taken into account in the 
comparison between theory and experiment. 

${\cal A}_{\Delta\Gamma}$ depends sensitively on NP and is hence
essentially unknown. Using (\ref{ys-LHCb}) and varying
${\cal A}_{\Delta\Gamma}\in[-1,+1]$ gives 
\begin{equation}
\Delta {\rm BR}(B_s \to \mu^+\mu^-)|_{y_s}=\pm y_s 
{\rm BR}(B_s \to \mu^+\mu^-)_{\rm exp},
\end{equation}
which has to be added to the experimental error of (\ref{defBrExp}).

In the SM, we have ${\cal A}_{\Delta\Gamma}^{\rm SM}=+1$ and rescale 
(\ref{BR-SM}) correspondingly by a factor of $1/(1-y_s)$, which results in 
\begin{equation}
\mbox{BR}(B_s\to \mu^+\mu^-)_{\rm SM}|_{y_s}=(3.5\pm0.2)\times 10^{-9},
\end{equation}
where we have used (\ref{ys-LHCb}). This is the SM reference  for
the comparison with the experimental branching ratio (\ref{defBrExp}). 

\boldmath
\section{The Effective $B_s\to\mu^+\mu^-$ Lifetime}
\unboldmath
With more data available, the decay time information can be included in the analysis.
As we pointed out in Ref.~\cite{BR-Bs}, the effective lifetime
\begin{equation}\label{effLifetime}
\tau_{\mu^+\mu^-} \equiv \frac{\int_0^\infty t\,\langle \Gamma(B_s(t)\to \mu^+\mu^-)\rangle\, dt}
	{\int_0^\infty \langle \Gamma(B_s(t)\to \mu^+\mu^-)\rangle\, dt}
\end{equation}
allows the extraction of 
\begin{equation}
 {\cal A}_{\Delta\Gamma} \, y_s = \frac{(1-y_s^2)\tau_{\mu^+\mu^-}-(1+
 y_s^2)\tau_{B_s}}{2\tau_{B_s}-(1-y_s^2)\tau_{\mu^+\mu^-}},
\end{equation} 
yielding
\begin{equation}\label{BR-correct}
\frac{{\rm BR}
	\left(B_s \to \mu^+\mu^-\right)}{{\rm BR}
	\left(B_s \to \mu^+\mu^-\right)_{\rm exp}}=2 - 
	\left(1-y_s^2\right)\frac{\tau_{\mu^+\mu^-}}{\tau_{B_s}}.
\end{equation}
We emphasize that it is crucial to the above equations that ${\cal A}_{\Delta\Gamma}$ 
in (\ref{ADG-lam}) indeed does not depend on the helicities of the muons, i.e. 
${\cal A}_{\Delta\Gamma}\equiv {\cal A}_{\Delta\Gamma}^\lambda$.

Effective lifetimes are experimentally accessible
through the decay time distributions of the same samples
of untagged events used for the branching fraction measurements,
as illustrated by recent measurements of the
$B_s^0\to J/\psi\,f_0$ and $B_s^0\to K^+K^-$ lifetimes~\cite{tauExp}
by the CDF and LHCb collaborations:
both attained a 7\% precision with approximately 500 events, while
an even larger sample of $B^0_s\to \mu^+\mu^-$ events can be collected
by the LHC experiments, assuming the Standard Model value of the
$B^0_s\to \mu^+\mu^-$ branching fraction.
Although a precise estimate is beyond the scope of this article,
we believe that the data samples that will be collected in the planned
high-luminosity upgrades of the CMS and LHCb experiments~\cite{Upgrade}
can lead to a precision of 5\% or better.
\boldmath
\section{Constraints on New Physics}
\unboldmath
In order to explore constraints on NP, we introduce
\begin{displaymath}
R\equiv 
\frac{\mbox{BR}(B_s \to \mu^+\mu^-)_{\rm exp}}{\mbox{BR}(B_s \to \mu^+\mu^-)_{\rm SM}}=
\left[\frac{1 + {\cal A}_{\Delta\Gamma} y_s}{1-y_s^2} \right]\left( |P|^2+ |S|^2\right)
\end{displaymath}
\vspace*{-0.6truecm}
\begin{equation}\label{R-def}
=\left[\frac{1+y_s\cos2\varphi_P}{1-y_s^2}  \right] |P|^2+
\left[\frac{1-y_s\cos2\varphi_S}{1-y_s^2}  \right] |S|^2,
\end{equation}
where we have used (\ref{ADG-lam}) and (\ref{BRratio}). Using (\ref{BR-SM}) and the 
upper bound \cite{LHCb-Bsmumu} yield $R<1.4$, neglecting the theoretical uncertainty from (\ref{BR-SM}).
In the case of $y_s=0$, $R$ fixes a circle in the $|P|$--$|S|$ plane. 
For non-zero $y_s$ values, $R$ gives ellipses dependent on the phases $\varphi_{P,S}$.
As these phases are in general unknown, a value of $R$ results in a circular band.
We obtain the upper bounds $|P|, |S|\leq\sqrt{(1+y_s)R}$. 
As $R$ does not allow us to separate the $S$ and $P$ contributions, there may still be a large 
amount of NP present, even if the measured branching ratio is close to the SM value.

\begin{figure}[!t]
    \centering
      \includegraphics[width=1.8in]{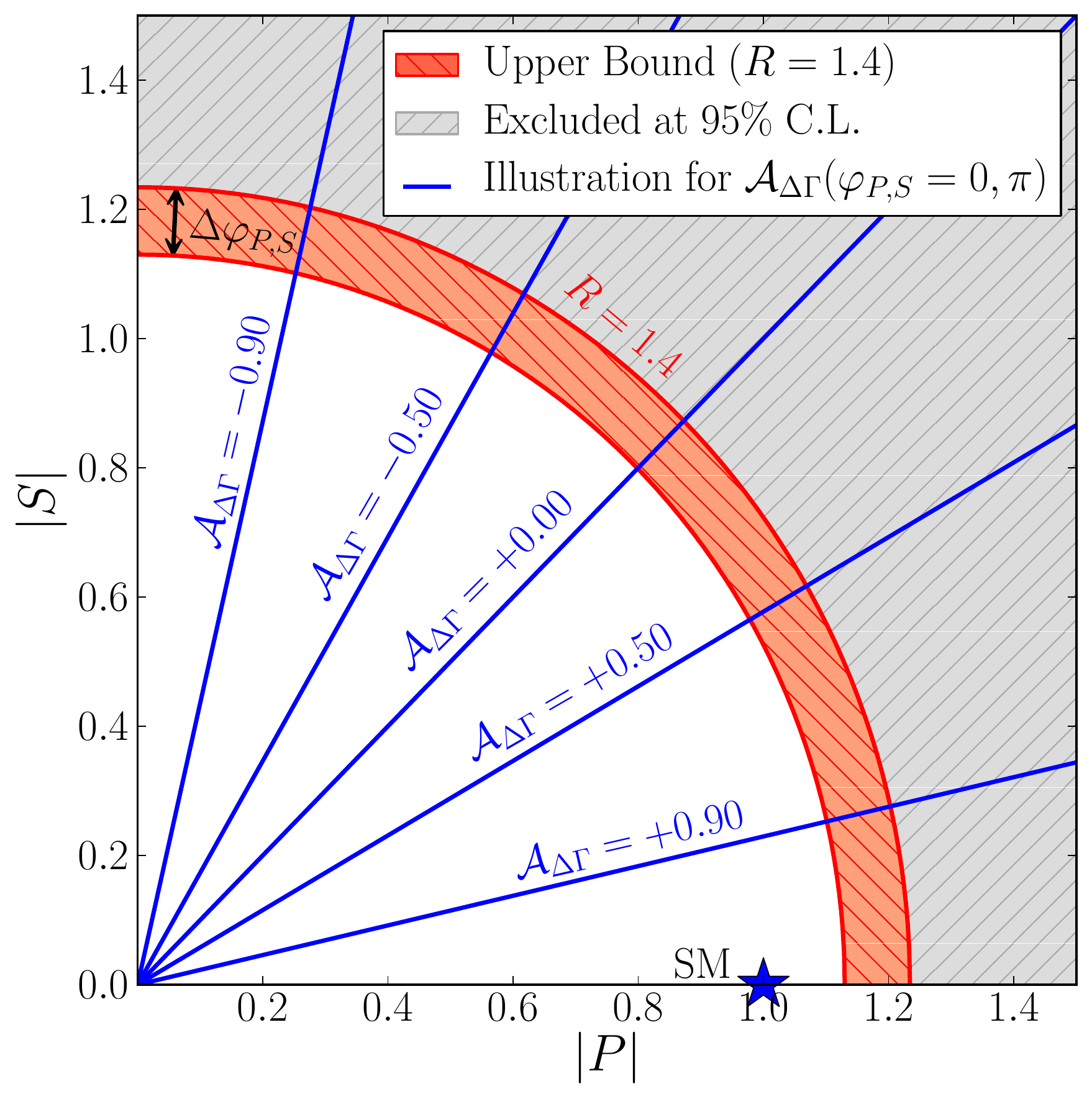} 
         \caption{\em Current constraints in the $|P|$--$|S|$ plane and illustration of 
    those following from a future measurement of the effective $B_s\to \mu^+\mu^-$ lifetime
    yielding the ${\cal A}_{\Delta\Gamma}$ observable.}\label{fig:fig-1}
 \end{figure}

\begin{figure}[!t]
    \centering
             \includegraphics[width=2.25in]{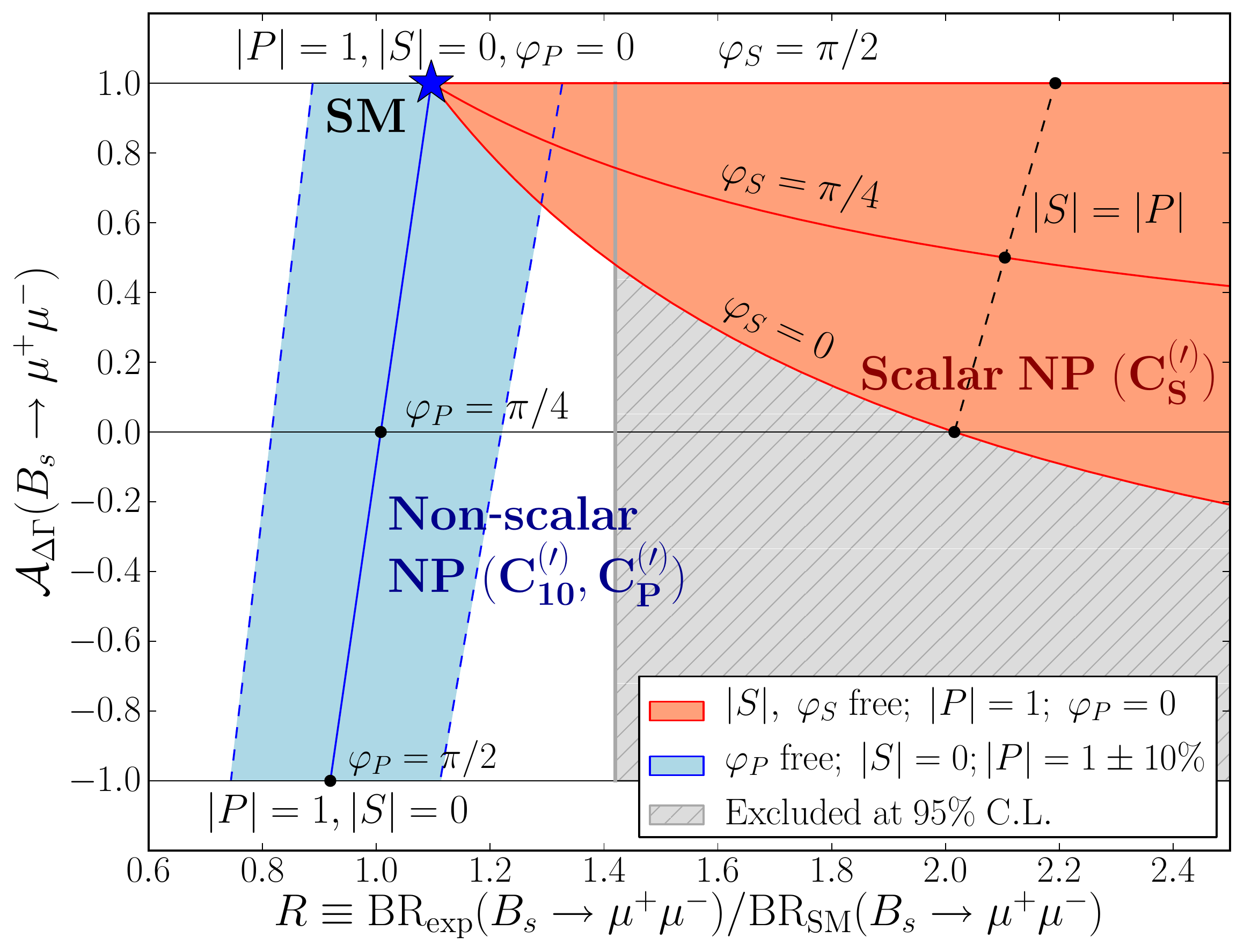} 
         \caption{\em Illustration of allowed regions in the $R$--${\cal A}_{\Delta\Gamma}$ plane for 
         scenarios with scalar or non-scalar NP contributions.}\label{fig:fig-2}
 \end{figure}

The measurement of $\tau_{\mu^+\mu^-}$ and the resulting observable
${\cal A}_{\Delta\Gamma}$ allows us to resolve this situation, as
\begin{equation}\label{S-ADG}
|S|=|P|\sqrt{\frac{\cos2\varphi_P-{\cal A}_{\Delta\Gamma}}{\cos2\varphi_S
+{\cal A}_{\Delta\Gamma}}}
\end{equation}
fixes a straight line through the origin in the $|P|$--$|S|$ plane. 
In Fig.~\ref{fig:fig-1}, we show the current $R$ constraints in the $|P|$--$|S|$ plane, and 
illustrate also those corresponding to (\ref{S-ADG}). In Fig.~\ref{fig:fig-2}, we 
illustrate the situation in the observable space of the $R$--${\cal A}_{\Delta\Gamma}$ plane. 
It will be interesting to complement these model-independent considerations with a
scan of popular specific NP models.

Let us finally note that the formalism discussed above can also straightforwardly be applied 
to $B_{s(d)}\to \tau^+\tau^-$ decays where the polarizations of the $\tau$ leptons can be 
inferred from their decay products \cite{CKWW}. This would allow an analysis of 
(\ref{asym-1}), where non-vanishing $C_\lambda$ observables would
unambiguously signal the presence of the scalar $S$ term. Unfortunately, these measurements
are currently out of reach from the experimental point of view.

\vspace*{-0.2truecm}

\section{Conclusions}
The recently established width difference $\Delta\Gamma_s$ implies that
the theoretical $B^0_s\to \mu^+\mu^-$
branching ratio in (\ref{BR-SM}) has to be rescaled by
$1/(1-y_s)$ for the comparison with the experimental branching ratio,
giving the SM reference value of $(3.5\pm0.2)\times 10^{-9}$.
The possibility of NP in the decay introduces an
additional relative uncertainty of $\pm 9\%$
originating from ${\cal A}_{\Delta\Gamma}\in[-1,+1]$.

The effective $B_s\to\mu^+\mu^-$ lifetime $\tau_{\mu^+\mu^-}$ offers
a new observable. 
On the one hand, it allows us to take into account the $B_s$ width
difference in the comparison between theory and experiments.
On the other hand, it also provides a new, theoretically clean probe of NP.
In particular, $\tau_{\mu^+\mu^-}$ may reveal large NP effects, 
especially those related to (pseudo-)scalar $\ell^+\ell^-$
densities of four-fermion operators originating from the physics beyond the SM,
even in the case that the $B^0_s\to \mu^+\mu^-$ branching ratio
is close to the SM prediction.

The determination of $\tau_{\mu^+\mu^-}$ appears feasible
with the large data samples that will be collected
in the high-luminosity running of the LHC with upgraded experiments
and should be further investigated, as this measurement
would open a new era for the exploration of $B_s\to \mu^+\mu^-$ at the LHC,
which may eventually allow the resolution of NP contributions
to one of the rarest weak decay processes that Nature has to offer.

\vspace*{-0.5truecm}

\section*{Acknowledgements} 
\vspace*{-0.3truecm}
This work is supported by the Netherlands Organisation for Scientific
Research (NWO) and the Foundation for Fundamental Research on Matter (FOM).


\end{document}